# Swarm Behavior of Intelligent Cloud

Anirban Kundu, Chunlin Ji

Kuang-Chi Institute of Advanced Technology
Shenzhen, Guangdong, P.R. China 518057
{anirban.kundu, chunlin.ji}@kuang-chi.org

## Abstract

In this paper, the main aim is to exhibit swarm intelligence power in cloud based scenario. Heterogeneous environment has been configured at server-side network of the whole cloud network. In the proposed system, different types of servers are being used to manage useful assorted atmosphere. Swarm intelligence has been adopted for enhancing the performance of overall system network. Specific location at server-side of the network is going to be selected by the swarm intelligence concept for accessing desired elements. Flexibility, robustness and self-organization, which are to be considered at the time of designing the system environment, are the main features of swarm intelligence.

## 1. Introduction

Swarm intelligence (SI) is the combined behavior of decentralized, self-organized systems, natural or artificial. The concept is engaged in effort on artificial intelligence. The expression was originally initiated by "Gerardo Beni" and "Jing Wang" in 1989, in the perspective of cellular robotic systems [1].

Swarm assumption has been built-up and discussed in [2]. Swarm behavior involves dynamism. SI systems are typically made up of a population of ordinary agents cooperating locally with one another and with their environment. The inspiration often comes from nature, especially biological systems. The agents follow very simple rules, and although there is no centralized control configuration dictating how individual agents should act, local, and to a certain degree arbitrary, communications among such agents direct to emergence of "intelligent" global behavior, unknown to the individual agents. Natural examples of SI include ant colonies, bird flocking, animal herding, bacterial growth, and fish schooling [3].

Group occurs at all magnitudes from microorganisms to whales, from factions of 10 to 10 million, and across a variety of sequential permanence from the transient congregation of midges to the mandatory regulations of herring. An aggregation may outline initially by arbitrary encounter and produce by density based connections. Group size is then determined by the balance of payoffs to individual members, where size of the group affects its performance [5].

Progression is over a much wider region and masses structure as much through contest between attractions as through confinement in tiny spaces in public festivals. Michael Batty et al. of University London developed a model to simulate the effect of changing the route of the Notting Hill Carnival in London. In the model the event space is first investigated by agents using ant scavenge symbol. Agents then shift in an unambiguous method to the event with information concerning space. Obstruction is progressively minimized commencing controls until a desired result is achieved. The consecutive stages of the simulation need interference by the management part of the event [6].

In this scenario, proposed approach shows the way to use swarm intelligence in cloud based systems for better dynamic formation of the server-side network.

Organization of rest of the paper is as follows: Section 2 presents related works section. Proposed approach is described in Section 3. Experimental results have been shown in Section 4. Section 5 concludes the paper.

## 2. Related Works

A system is a collection of machines, workstations, servers and some other resources connected by networks. Distributed performance computing [7] in heterogeneous systems employs the distributed objects as applications. These applications are arranged in such a manner that the same type of user requests can be executed in distinct machines which are situated in different locations (in case of Wide Area Network (WAN)). Sometimes, these machines fall in the same group or cluster at same location (in case of Local Area Network (LAN)) [8]. Formation of distributed network requires transferring information in a high speed [9]. Thus, compiler mapping has to be done for communication sensitive jobs in different server machines of the network. Remote Procedure Call is the most well-known method for machine to machine communication handshaking for interconnecting heterogeneous systems. Sometimes, Remote Method Invocation (RMI) is also used within the network. In real-time scenario, there may be several instances when the server assignments are not appropriate. Server assignment problem can be occurred due to heterogeneous distributed environment. Different types of scheduling mechanisms are classified based on the requirements. In heterogeneous system environment, several types of servers can be implemented for better selection of tasks to be compiled and executed for heterogeneous supercomputing. High performance can be achieved using cluster of workstations [10-11]. These workstations can be of similar types or distinct types in respect of configurations. Therefore, practical situation of the distributed network needs multi-processing [12] power within the network activities to speed-up the task with synchronization [13]. Whenever a transaction is taking place in a server-side machine for a particular process, all other processes would wait till its completion [14]. Another important point is transactional memory and garbage collection comparison which is dealt in [15]. In distributed systems, shared memory concept should be applied for multi-processor operating systems. In this way, the network memory can be considered as local memory [16]. Proper management of central processing unit, memory and network bandwidth is required while the external users access server-side resources having limitations [17]. Concurrent execution of procedures has been accomplished in the network scenario maintaining specific data structures [18].

## 3. Proposed Approach

In cloud network, typically a lot of servers are being utilized to accommodate all the users' requests in real-time basis. So, a huge data and/or information have to be transferred from one location to another location for execution of several programs based on the requirement of physical memory, virtual memory, disk space, and so on. Some common features are there in both swarm intelligence and in cloud system. So, a cloud system can be configured like a swarm based activity. Cloud system network has some interesting behaviors like servers are clustered as sub-network within the network. The structure can be created following specific sorting mechanisms. Cooperative transport is one of the major points of consideration in cloud based system. Otherwise, the synchronization would not be happened properly. The whole hierarchical structure of servers and workstations should be maintained following some protocols. These points are also true for swarm intelligence. Cloud system typically has two types of interactions such as direct and indirect. Peer-to-peer data transfer is accomplished by direct interactions whereas monitoring system handles all the situations in the network using indirect interactions.

Sometimes, individual system behavior modifies the cloud environment, which in turn modifies the behavior of other individuals of the system. This can be called as local-to-global and global-to-local transition. This indirect interaction is known as "Stigmergy" in swarm intelligence. Intelligent group activities are highly required for maintaining synchronization between all the server machines within the network. High computations are typically handled by the high-end cluster machines whereas small computations are handled by the low processors. These types of distinction of tasks are clearly visible in swarm activities as the division of labor and adaptive task allocation. In the proposed approach, nearest network and the nearest server within that network have been utilized to execute the client based programs using the similarity concepts. This similarity concept is similar to the discovery of the shortest paths between the source and the destination. Overall, the collective behavior of swarm intelligence is visible in the proposed cloud based system network which is described later in this section with illustrations. In this approach, feedback mechanism is used for controlling any abnormalities. It is called as load shifting concept within this paper. Random behavior in multiple interactions is also common in proposed cloud system and swarm system (refer Figure 1).

Distributed environment in a heterogeneous system has become a smart alternative for distributing high performance on a variety of system defined and/or used defined functions. Disseminated control processing has been supported on progresses in knowledge and expertise areas high performance arrangements. Existing application packages enable network based distributed processing, task multi-programming, and compilation techniques for distributed-memory multiple instruction multiple data (MIMD) computers. The paper proposes a Software-as-a-Service (SaaS) providing high services and less maintenance to the users. It enables centralized control of business (Simulation software) by the service provider. SaaS is basically a technique in demand of the software users. Its on-demand licensing has unfasten the cost of providing every machine with the service that is used only when needed

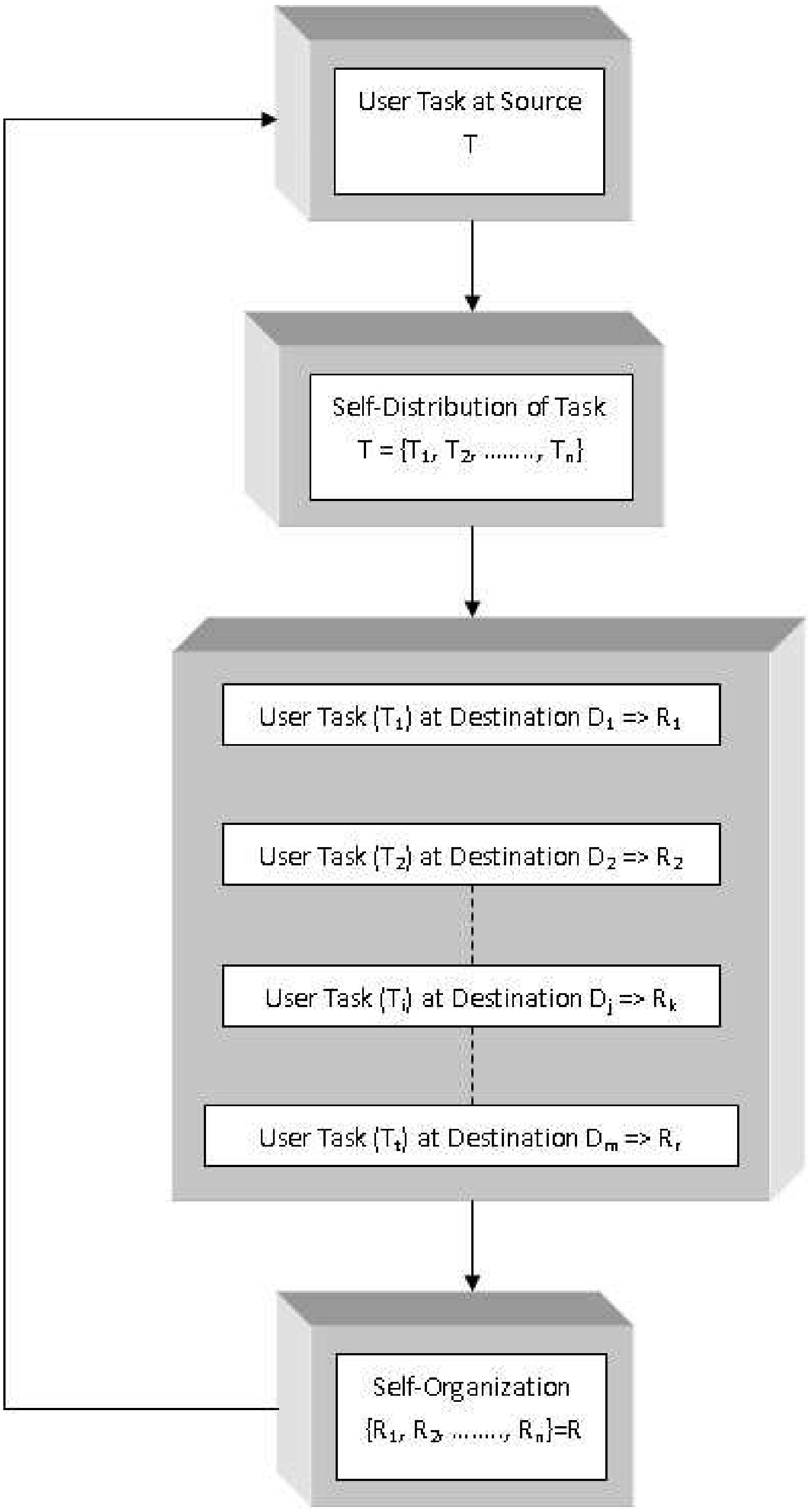

Figure 1: Implementation of Self-Organization Concept in Cloud System Network

In this paper, a distributed environment has been measured as exposed in Figure 2. It consists of typical computer resources, workstations and clusters. This gathering of equipments presents a huge widespread computing resource including memory, cycles, storage and bandwidth. Proposed system has a great prospective for high performance computing in this approach. An important characteristic of this structural arrangement is that it exhibits heterogeneity of many types including hardware, operating system, file system, and network. Heterogeneity creates a challenge that it must be managed to enable the parts of the proposed system to work together. At the same time, it also presents an opportunity that is the variety of different resources which suggests that it is possible to select the best resources for a particular user request. The variety and amount of computing resources in the proposed system offers a great prospective for high performance computing. In Figure 2, each network ($N_1$, $N_2$, ……., $N_i$, ……., $N_n$) is connected using routers/switches. Several numbers of servers having different standards and configurations are directly or indirectly connected to these routers at network level.

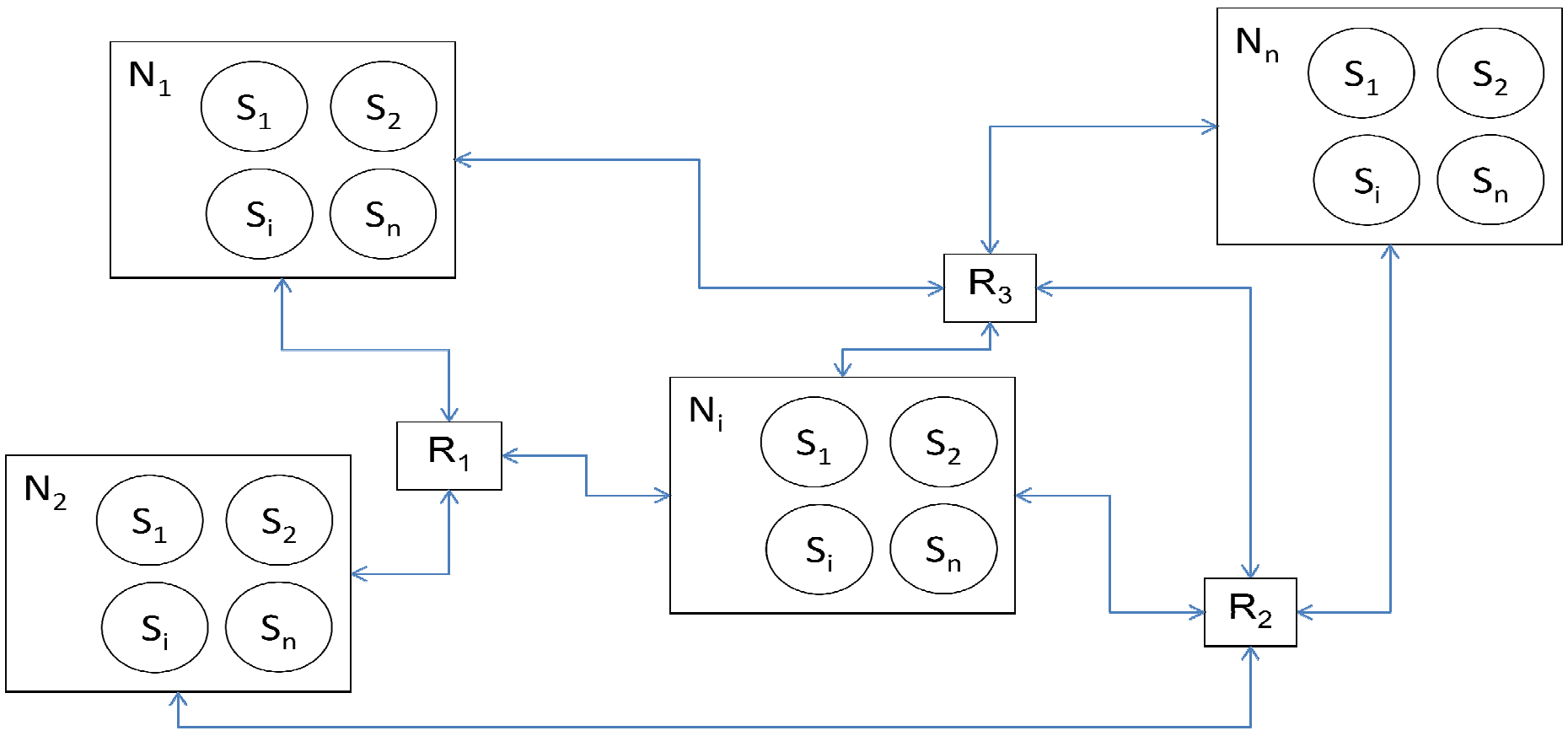

Figure 2: Intra-Network Structure of proposed approach

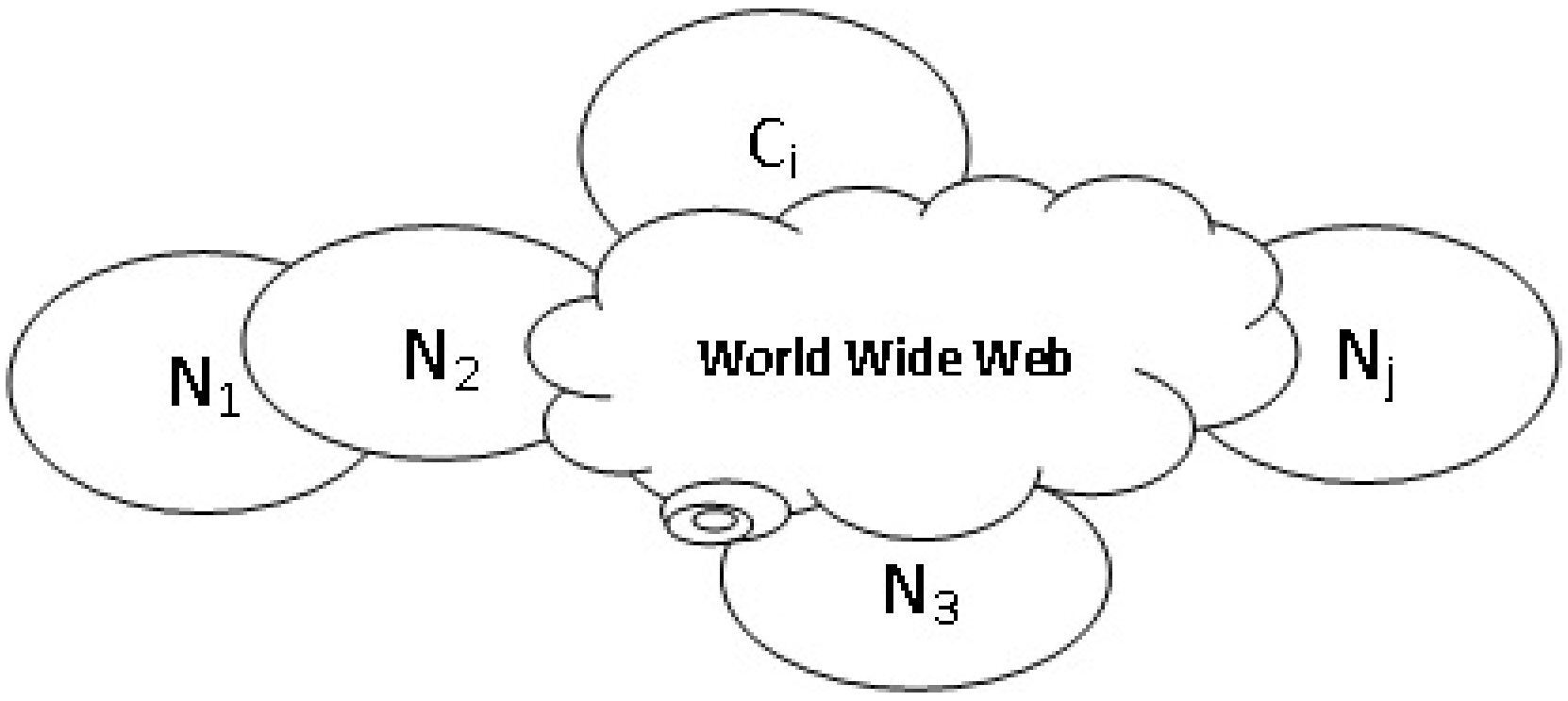

Figure 3: Network Connectivity between different server networks and Clients using World Wide Web

$$\text{SimilarityN}\big(N(i), N(j)\big) = \frac{\text{AoAN}(\text{AppN}(N(i)) \cap \text{AppN}(N(j)))}{\text{AoAN}(\text{AppN}(N(i)) \cup \text{AppN}(N(j)))} \qquad (1)$$

where, N(i) = $i^{th}$ network;

N(j) = $j^{th}$ network;

AppN() = Function of "Number of Applications of individual network";

AoAN() = Function of "Amount of Applications of considered networks";

SimilarityN() = Function of "Similarity Measurement between networks";

$$\text{SimilarityS}\big(S(i), S(j)\big) = \frac{\text{AoAS}(\text{AppS}(S(i)) \cap \text{AppS}(S(j)))}{\text{AoAS}(\text{AppS}(S(i)) \cup \text{AppS}(S(j)))} \qquad (2)$$

where, S(i) = $i^{th}$ server;

S(j) = $j^{th}$ server;

AppS() = Function of "Number of Applications of individual server";

AoAS() = Function of "Amount of Applications of considered servers";

SimilarityS() = Function of "Similarity Measurement between servers";

“SimilarityN” and “SimilarityS” are defined in Equation 1 and Equation 2 respectively to facilitate the nearest network and nearest server concept. Related networks would be responsible to execute the user specific tasks. Network manager directs the user initiated information and data to the nearest network for carrying out assignment. Related server is detected to process the user request in an actual manner as a service provider. After completion of the whole procedure, output information would be sent to the user-end following the counter way. Server $S_k$ of Network $N_j$ detects the geographical location of $C_i$ using user IP as depicted in Figure 3.

Figure 4 describes the stage-wise execution of different parts of the proposed system. Here, the whole system behaves like a SaaS aiming to provide a full fledged implementation of network based activities along with the dynamic scalability, openness, distributive nature and several transparencies. It aims to give the customer a unique feel like “every computer is your personal computer”. Software should be customized according to the specific users. Requirement of users vary. At present market, users have to buy full package of any software. But in our approach, users are free to decide which module they wish to use; i.e., customers can choose any or all the modules according to their needs.

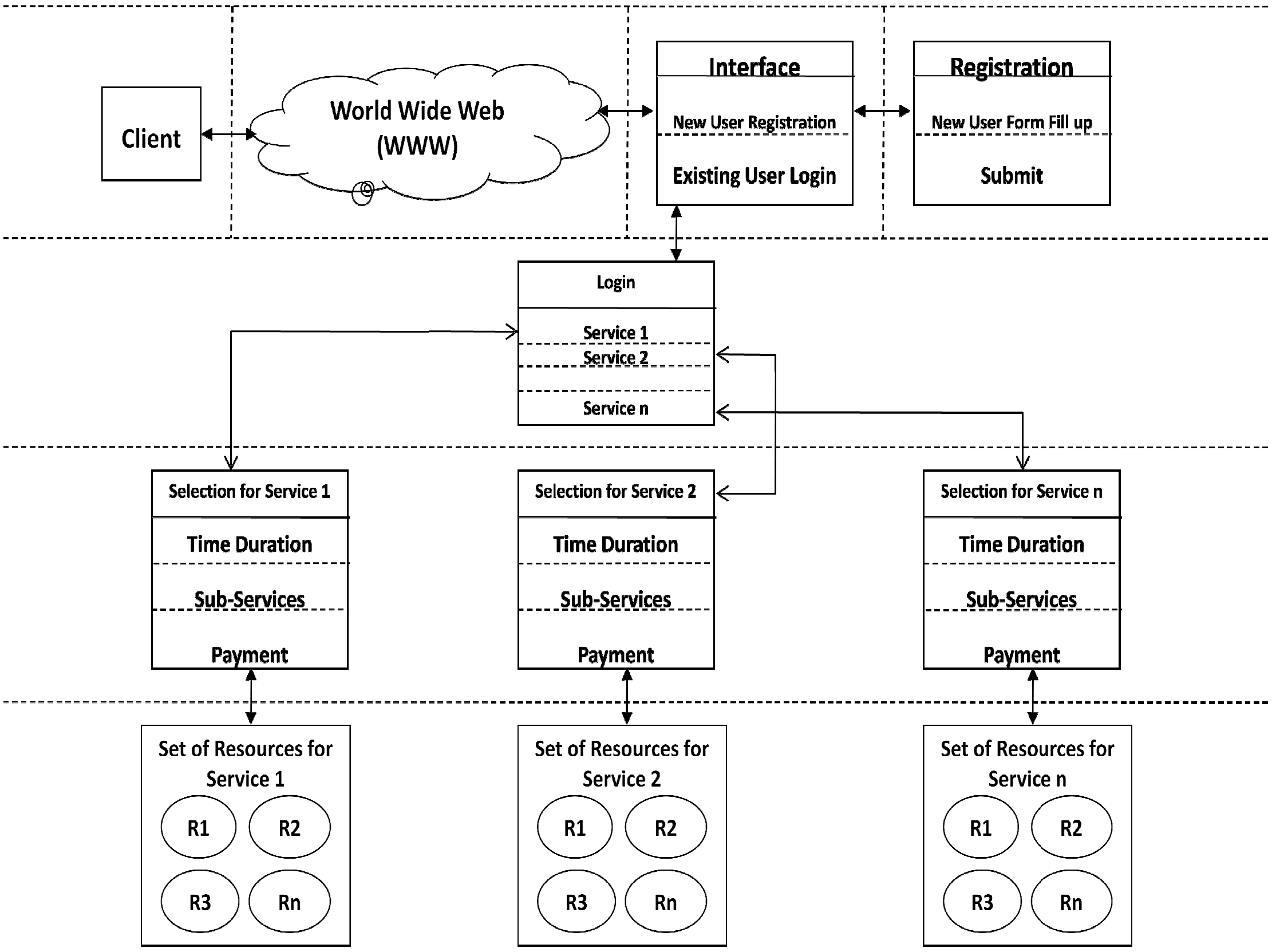

Figure 4: Different levels of proposed framework

It has been shown that dynamic scalability is quite possible in the network activities if the proposed design is perfect for it. Designers should provide some specific methodologies for managing flawless activities in dynamic scalability in real-time. Algorithm 1 explains the overall cloud activity for specific applications in Web services.

Algorithm 1: Cloud Activity in Web Services

Input: Request from user (user application, user data with time stamp, user IP)
Output: Network selection for program execution of user

Step 1: Each server maintains the information of all other servers within a specific network. A network manager maintains the information about all other related networks within the Internet.
Step 2: When a client requests for data, the information related to the user is broadcasted on the Web.
Step 3: Determine nearest network ($N_j$)
Step 4: $N_j$ detects the user request and accepts it.
Step 5: Specific server ($S_k$) is appointed for handling client requests ($C_i$).
Step 6: After the transaction is over, $S_k$ sends the update of $C_i$ to each related server of $N_j$. Therefore, the information is propagated to $\sum_{k=1}^{n}$ Sk. Consider, total number of related servers within $N_j$ = n.
Step 7: Calculate access frequency of the user request for particular application by comparing with other applications
Step 8: Network $N_j$ transmits the update to other related networks.
Step 9: Each related network updates the related servers.
Step 10: Stop.

Nearest network has been determined using a seed which has been chosen based on user. Another seed is chosen based on the IP address of the user location. Hash function is selected considering the characteristics of these seeds. Particular network is selected comparing the current load with the threshold load of the network (refer Figure 5).

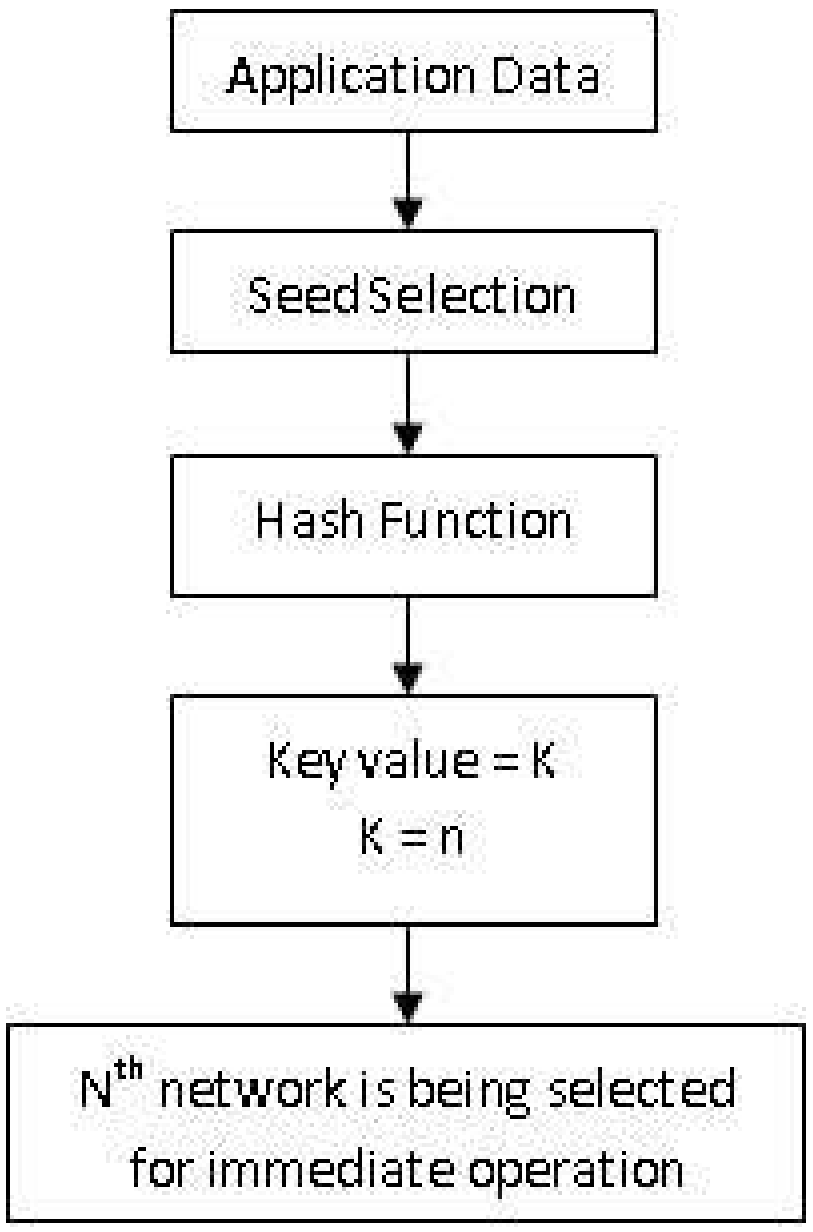

Figure 5: Flowchart for network selection

There are mainly two points which have to be verified before sending the actual information. Higher access frequency is considered first for transmission over the network. Old information should be sent earlier in case of having same access frequencies. In similar way, information is transmitted to related servers after it is being reached to the specific network.

Initially at server-side, the document for client's Internet Protocol (IP) address has been stored by the Web portal system using TCP/IP connection and then further submitted to the interface layer to activate the proposed distributed system architecture. The user defined document for the input data for a specific program/activity should also be saved in the same procedure. I/P sender module takes care of these documents and prepare the required formatted materials for next level of processing to search for a particular sub-network within the server-side system based on the classified tasks using dynamic scheduling technique. Scheduler checks for an active server of a particular network at first, and then it checks whether the server is busy or free. The particular server is being selected for the user prescribed operation if the server is active and free at any time instance. Otherwise, another server would be selected for the same operation. I/P sender module also send the required materials to I/P receiver module following the standardized typical communication techniques at port level. I/P receiver module initiate the particular application. The output files would be stored in a particular location of each server machines. O/P sender module collects the data and sends it back to the interface. O/P receiver module fetches the required output files coming to its direction at a particular port level and stores the required information to the prescribed places related to the external IP address of the particular user. Web portal takes care of these generated documents and sends them to the specific destination through TCP/IP connection (refer Figure 6).

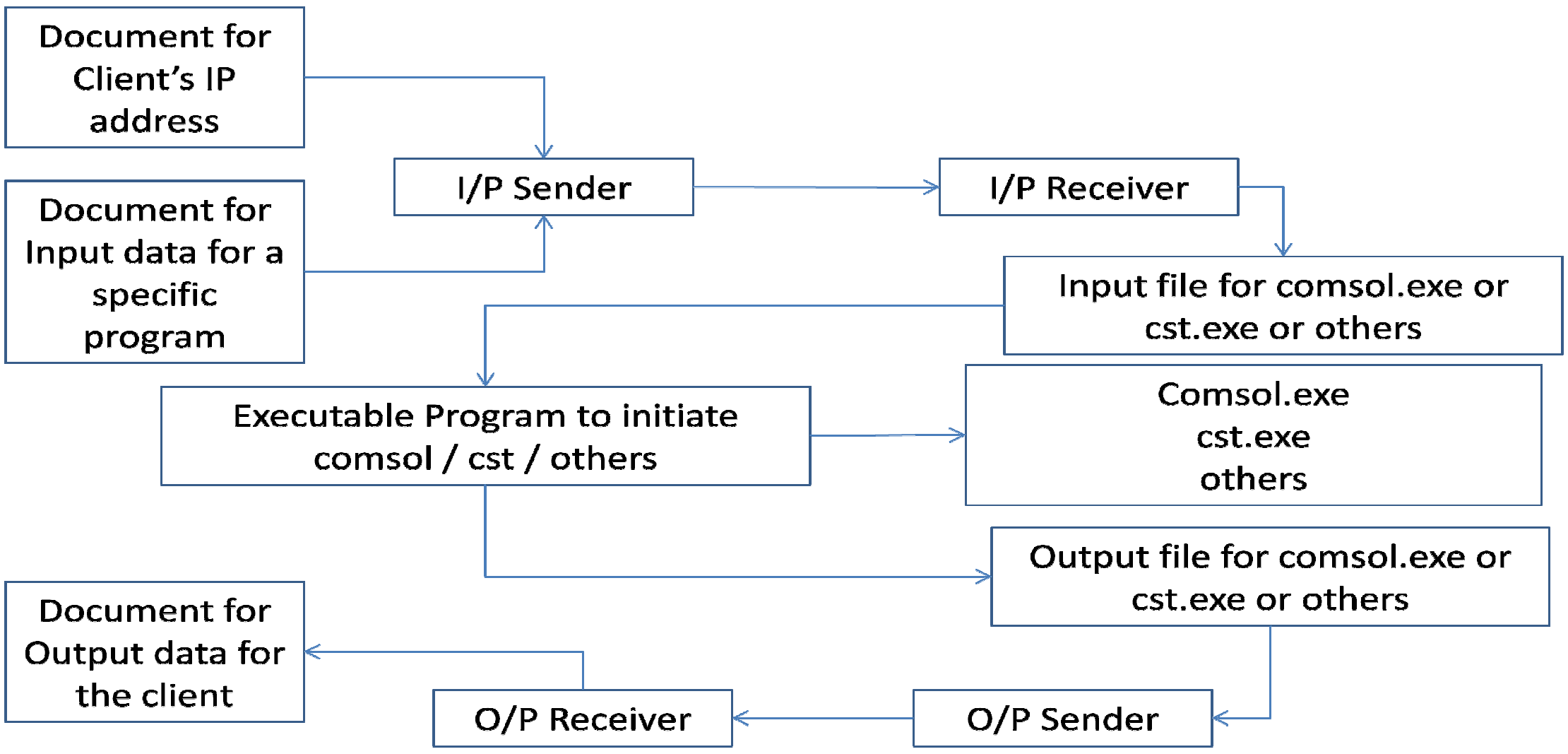

Figure 6: Interaction between Sender & Receiver

Table 1: Network-Server Map (NS Map)

| I | $S_1$ | $S_2$ | $S_3$ | ……….. | $S_j$ | ……….. | $S_s$ |
|---|---|---|---|---|---|---|---|
| $N_1$ | $App_1$, $App_2$, $App_3$, $App_4$ | $App_1$, $App_2$, $App_4$ | $App_1$, $App_4$ | ……….. | $App_2$, $App_3$ | ……….. | ……….. |
| $N_2$ | $App_1$ | $App_1$, $App_2$ | $App_1$, $App_2$, $App_3$ | ……….. | $App_3$ | ……….. | ……….. |
| $N_3$ | $App_1$ | $App_1$ | $App_1$, $App_3$ | ………… | $App_2$ | ………… | ………… |
| $N_4$ | $App_2$, $App_3$ | $App_3$, $App_4$ | $App_2$, $App_4$ | ………… | $App_2$, $App_3$, $App_4$ | ………… | ………… |
| ●<br>●<br>● | ●<br>●<br>● | ●<br>●<br>● | ●<br>●<br>● | ●<br>●<br>● | ●<br>●<br>● | ●<br>●<br>● | ●<br>●<br>● |
| $N_i$ | ………… | ………… | ………… | ………… | ………… | ………… | ………… |
| ●<br>●<br>● | ●<br>●<br>● | ●<br>●<br>● | ●<br>●<br>● | ●<br>●<br>● | ●<br>●<br>● | ●<br>●<br>● | ●<br>●<br>● |
| $N_n$ | ………… | ………… | ………… | ………… | ………… | ………… | ………… |

Table 1 shows the mapping between networks and servers within each network. It is treated as the hierarchical formation of proposed design. 'n' number of networks have been considered and 's' number of servers have been considered in each network. Each server consists of some applications as predefined by the network system administrator.

Here, Interface is denoted by I.
Network is denoted by N.
Server is denoted by S.
Consider, number of networks = n
Number of servers within a network = s
Therefore, $N_i$ is the $i^{th}$ network where i = {1, 2, 3, ……….., n}
Therefore, $S_j$ is the $j^{th}$ server of a particular network where j = {1, 2, 3, ………………, s}
User request can be treated as $U_x$. 'x' means the application name such as "comsol", "matlab", "cst", etc..
Interface of the whole network generates the function as $F_I(U_x)$.
Apply $F_I(U_x)$ to available networks as follows:
$F_I(U_x) \rightarrow \{N_1, N_2, \ldots, N_i, \ldots, N_n\} = \{F_{N1}(F_I(U_x)), F_{N2}(F_I(U_x)), F_{Ni}(F_I(U_x)), F_{Nn}(F_I(U_x))\}$
Therefore, $F_I(U_x) \rightarrow \int_{i=1}^{n} N_i = \int_{i=1}^{n} F_{N_i}(F_I(U_x))$
Now, apply each $F_{Ni}(F_I(U_x))$ to all the servers of each network ($N_i$) as follows:
$F_I(U_x) \rightarrow \{N_1 \rightarrow \{S_1, S_2, \ldots, S_i, \ldots, S_s\},\ N_2 \rightarrow \{S_1, S_2, \ldots, S_i, \ldots, S_s\},\ N_i \rightarrow \{S_1, S_2, \ldots, S_i, \ldots, S_s\},\ N_n \rightarrow \{S_1, S_2, \ldots, S_i, \ldots, S_s\}\} =$
$\{F_{S1}(F_{N1}(F_I(U_x))), F_{S2}(F_{N1}(F_I(U_x))), \ldots, F_{Si}(F_{N1}(F_I(U_x))), \ldots, F_{Ss}(F_{N1}(F_I(U_x)))\}$,
$\{F_{S1}(F_{N2}(F_I(U_x))), F_{S2}(F_{N2}(F_I(U_x))), \ldots, F_{Si}(F_{N2}(F_I(U_x))), \ldots, F_{Ss}(F_{N2}(F_I(U_x)))\}$,
…….…….…….…….,
$\{F_{S1}(F_{Ni}(F_I(U_x))), F_{S2}(F_{Ni}(F_I(U_x))), \ldots, F_{Si}(F_{Ni}(F_I(U_x))), \ldots, F_{Ss}(F_{Ni}(F_I(U_x)))\}$,
…….……….…….,
$\{F_{S1}(F_{Nn}(F_I(U_x))), F_{S2}(F_{Nn}(F_I(U_x))), \ldots, F_{Si}(F_{Nn}(F_I(U_x))), \ldots, F_{Ss}(F_{Nn}(F_I(U_x)))\}$

Therefore, $F_I(U_x) \rightarrow \int_{i=1}^{n} N_i \rightarrow \int_{j=1}^{s} S_j = \int_{j=1}^{s} F_{S_j} \int_{i=1}^{n} F_{N_i} \left(F_I(U_x)\right)$

Proposed system framework detects the nearest network and server for a particular application at any specific time instance based on the user request $U_x$.

There is one option for handling emergency situation of system crash. It is known as load shift. The required output would not be received by the interface, if the server system crashes down. Therefore, the whole system network needs a supervision to tackle this type of scenario. A special monitoring sub-system at interface has been settled as a result. This monitor checks about the status of the particular network as well as the particular server(s) of that network. If the monitor detects any unusual activity within the network and/or server, it would execute the same procedure in some other servers of the same or different network. If there is no response from the network and/or server, then also the interface does the same procedure in an alternate network and/or server. Algorithm 2 describes the load shifting technique as follows:

Algorithm 2: Emergency_Load_Shift

Input: Total list for a particular task, T = $\int_{i=1}^{n} T_i$ = {$T_1$, $T_2$, ……………, $T_i$, ……………, $T_n$}
Output: Execute missing links to complete task

Step 1: Wait for 't' time units
Step 2: If (All links have reached interface)
Then goto Step 5
Step 3: Else
Find missing links of T = $\int_{i=1}^{n} T_i$ (Call Algorithm 3)
Step 4: Execute missing links in different servers (Sj) of sub-network ($N_i$)
Step 5: Stop

Algorithm 3 has been invoked for handling the real-time log structure of the system, when missing links are being tracked by Algorithm 2. Table 2 shows a sample log file of the system. It contains external IP information, the application name, internal IP information and the number of files required as output of the prescribed program module.

Table 2: Structure of a Log File

| LOG FILE | | | |
|---|---|---|---|
| External IP | Application | Internal IP (Server) | No. of Files |
| 10.20.30.40 | App1 | 192.168.10.50 | 4 |
| 10.20.30.41 | App2 | 192.168.10.80 | 6 |

Algorithm 3 describes the detailing about the missing links. Here, missing link means the information about the untraceable particular program request to a specific server of a specific sub-network. To accomplish this task, the request has to be executed again in some other server machines of the network. The network and the corresponding server(s) are being assigned in run-time.

Algorithm 3: Find_Missing_Links

Input: Log file

Output: Fix missing links

```
Time_Interval_1 = 1000 (1 second)
Time_Interval_2 = 300000 (5 minute)
While (TRUE)
{
    Check (Log_File)
        If (Open)
            Wait (Time_Interval_1)
            Continue
        Else If (! Open)
            Break

    Open (Log_File)
    Fetch (ith Line)

    Check (Database_External_IP)
        If (! Found)
            Delete the entry from log_file
            Goto next entry of External IP
        If (Found)
            Break

    Check (Database_External_IP for the exact no. of files)
        If (Found) // no. of files
            Delete the entry from log_file
            Goto next entry of External IP
        Else If (! Found)
            Check the connection between Interface and Server
            If (Connection)
                Goto next entry of External IP
            Else If (! Connection)
                Check (Active_List)
                    If (! Found)
                        Delete the entry from log_file
                        Goto next entry of External IP
                    Else If (Found)
                        Check (Busy_List)
                            If (! Busy)
                                Delete (Entry_from_Active_List)
                                Delete (Entry from Log_File)
                                Goto next entry of External IP
                            Else If (Busy)
                                Delete(Entry_from_Busy_List)
```

```
                        Delete (Entry_from_Active_List)
                        Emergency (Signal_to_system)
                        Goto next entry of External IP

    Close (Log_File)
    Wait(Time_Interval_2)
}
```

## 4. Experimental Results

All stages of proposed approach have been illustrated in this section using the relevant examples. The user registers through the available Web gateway. Then, user tenders specific individual information within the required online forms. Technical information is also being submitted to the organizational database for further communication. After that, user is competent to sign-in within the proposed system for allocation of resources.

Figure 7 demonstrates dynamic activity of IP_SEND module for controlling initial client data preserving organization between the interface and other related server machines of different internal networks of the disseminated setting. All the essential information has been thrown to the explicit servers for implementation. IP address of the outside client has to be sent to the particular server of the nearest network. It would be necessary to switch the fractional productions of the user's program.

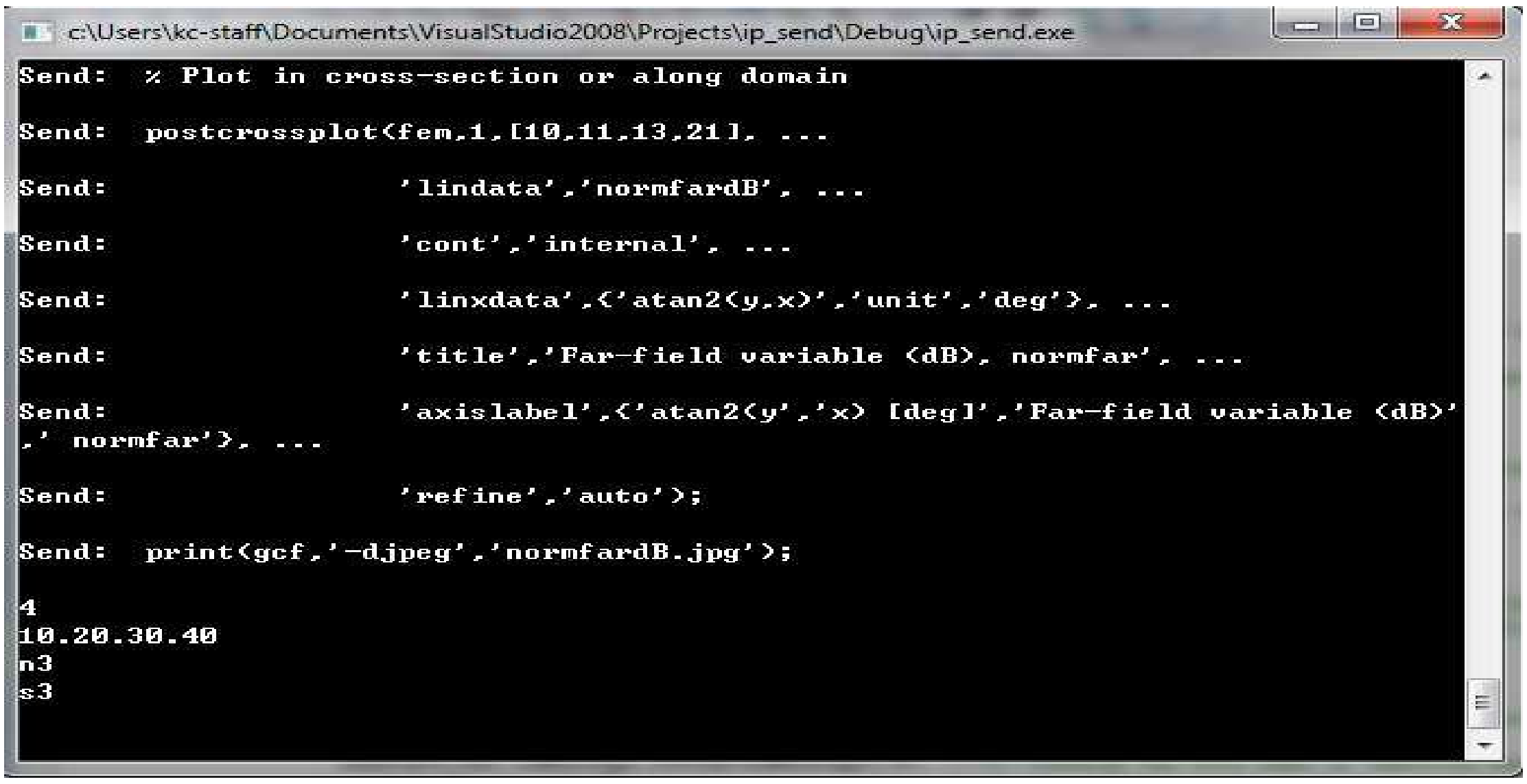

Figure 7: Presentation of IP_SEND Module

Figure 8 depicts IP_RECEIVE unit which is accountable to carry out actual implementation of the client presented agenda. This module first obtains the agenda after acknowledgement by interface as a associated server of the nearest network. It receives output type along with server IP, nearest network and nearest server. It creates a result file irrespective of the output types. The downloaded file(s) would be stored within predefined repository of the server in a provisional mode. Then, the module identifies the meticulous executables at kernel level.

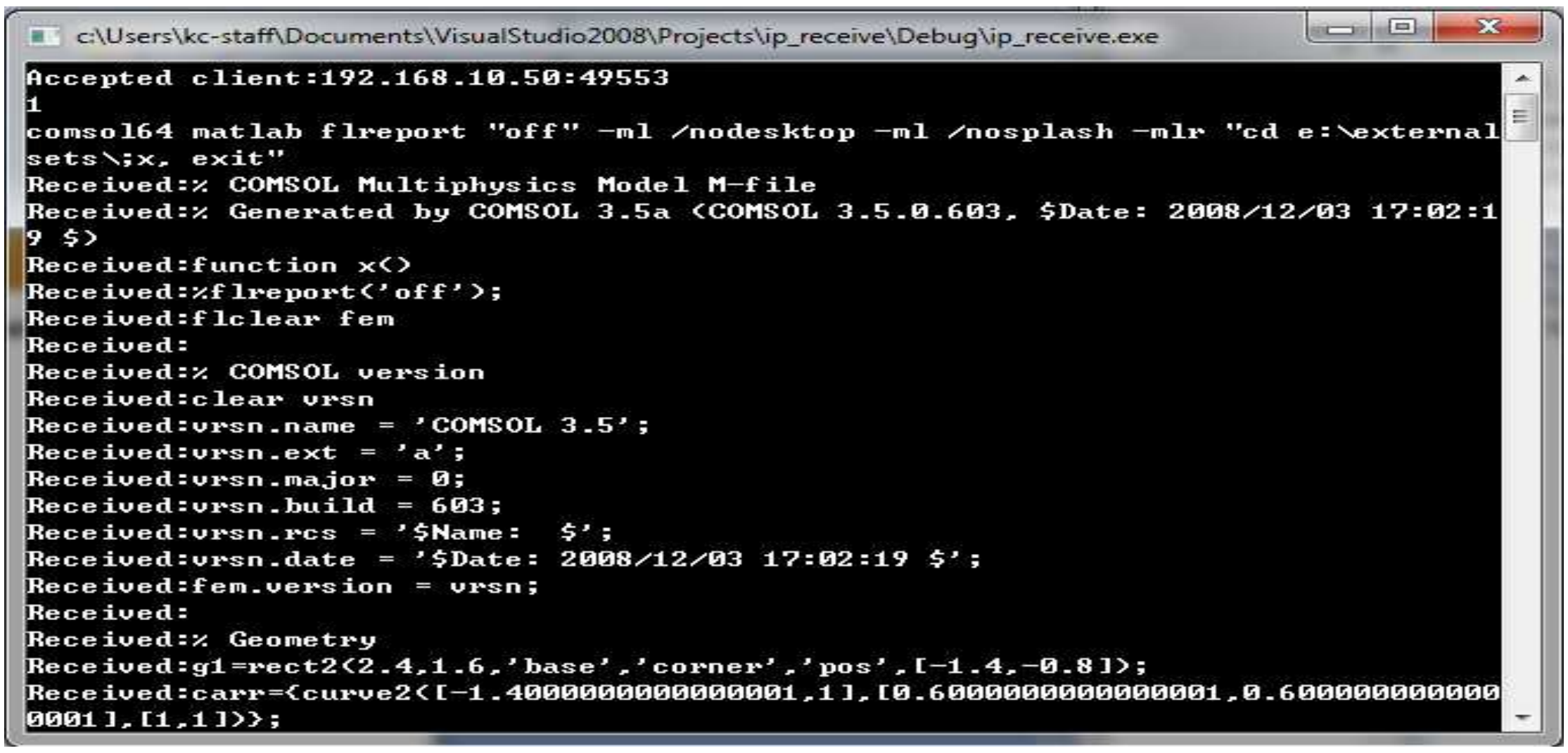

Figure 8: Presentation of IP_RECEIVE Module

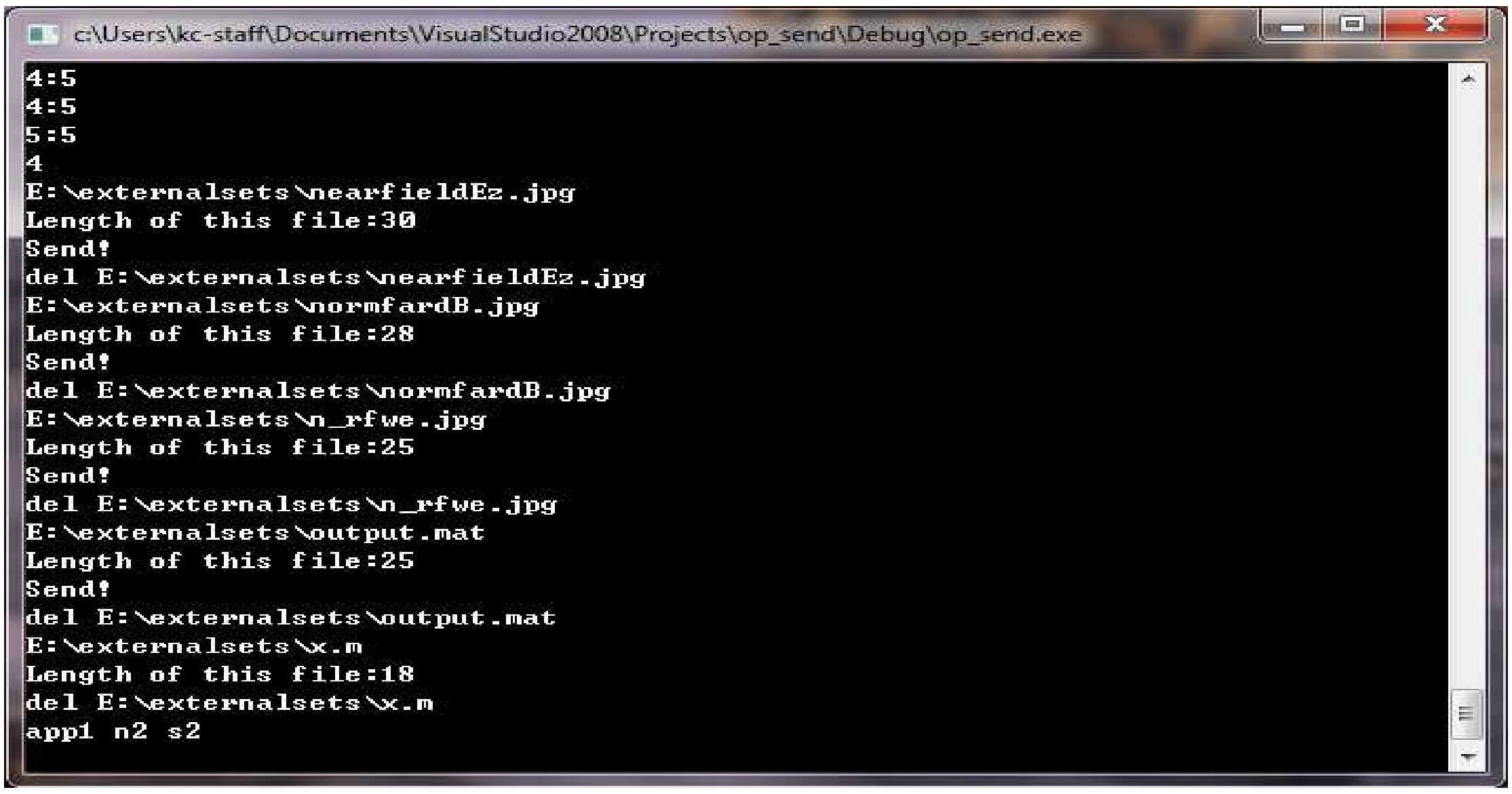

Figure 9: Presentation of OP_SEND Module

Figure 9 symbolizes an example of real-time presentation of OP_SEND module. This unit receives the outputs of the server-side series of coding preferred by the external client and then relocates these records to the interface. The module transmits the required output information after completion of execution.

Figure 10 represents the performance of OP_RECEIVE module at interface level. It receives all the outputs and stores the documents inside selected index for the meticulous client. This module helps to control the busy information about the concerned networks and their servers. The specific server is being released after its execution. Therefore, the server can be consumed for other tasks in upcoming events.

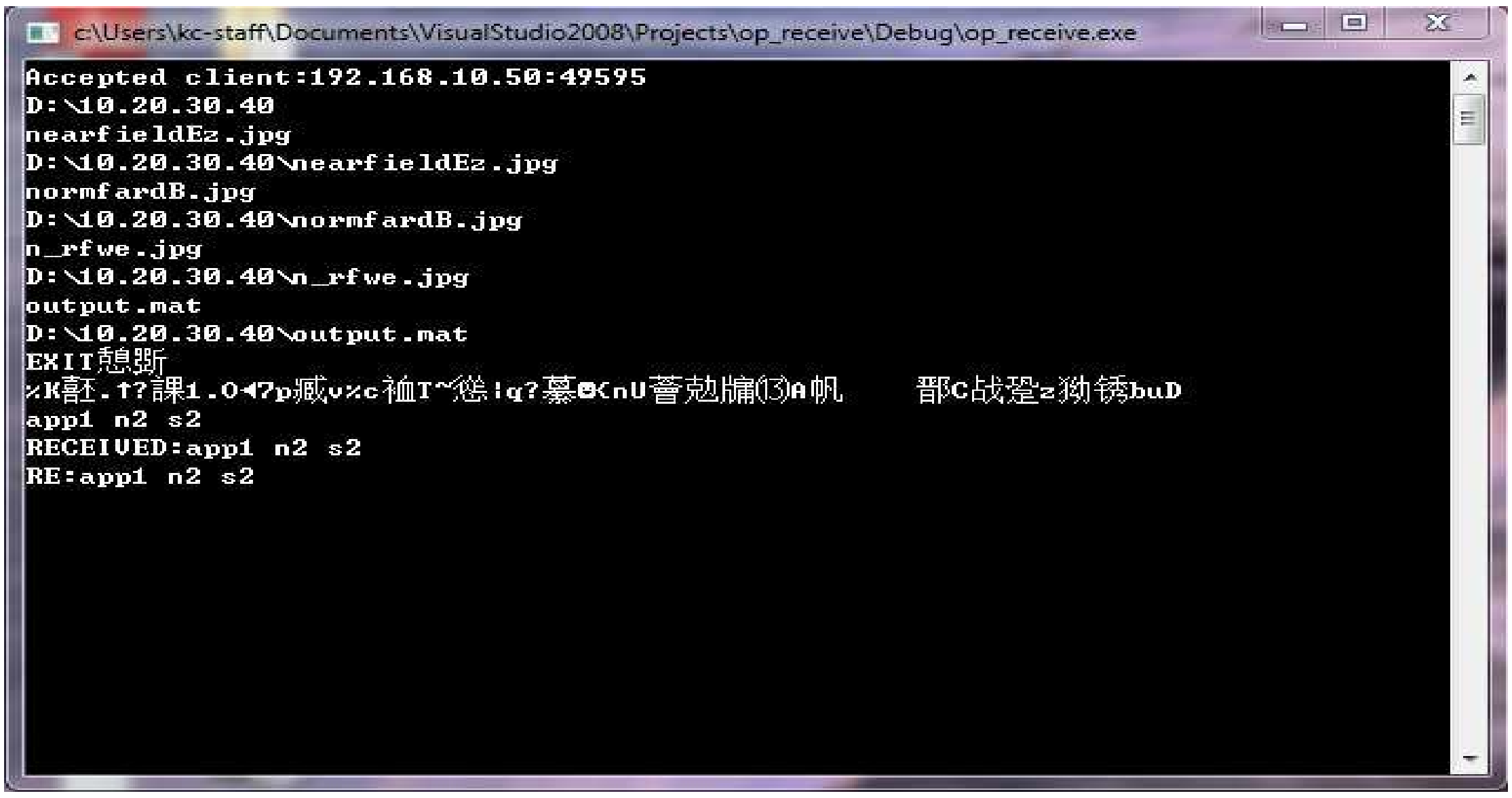

Figure 10: Presentation of OP_RECEIVE Module

## 5. Conclusion

In this paper, distributed environment has been achieved using swarm intelligence. Synchronized system network has been formed with different types of servers. “Stigmergy” is successfully applied to the proposed system for enhancing swarm concept. Self-organizing behavior is implemented artificially in the cloud to prepare and control the system as swarm movement in real-time. “COMSOL” and “MATLAB” are considered for the case studies and experiments. Any type of software can be implemented in this environment for better efficiency in respect of time and speed. Flexibility, robustness and self-organization have been considered at the time of designing the system environment.